\documentclass[nofootinbib,aps,prl,reprint,superscriptaddress]{revtex4-2}

\usepackage{graphicx}
\usepackage{dcolumn}
\usepackage{bm}
\usepackage{xcolor}
\begin{document}


\title{Laser spectroscopy of neutron-rich $^{207,208}$Hg isotopes: Illuminating the kink and odd-even staggering in charge radii across the $N=126$ shell closure}



\author{T.~Day~Goodacre}
\email{tdaygoodacre@triumf.ca}
\affiliation{The University of Manchester, School of Physics and Astronomy, Oxford Road, M13 9PL Manchester, United Kingdom}
\affiliation{CERN, CH-1211 Geneva 23, Switzerland}
\affiliation{TRIUMF, Vancouver V6T 2A3, Canada}

\author {A.V.~Afanasjev}
\affiliation{Department of Physics and Astronomy, Mississippi State University, MS 39762, USA}

\author {A.E.~Barzakh}
\affiliation{Petersburg Nuclear Physics Institute, NRC Kurchatov Institute, Gatchina 188300, Russia}

\author {B.A.~Marsh}
\affiliation{CERN, CH-1211 Geneva 23, Switzerland}

\author{S.~Sels}
\affiliation{CERN, CH-1211 Geneva 23, Switzerland}
\affiliation{KU~Leuven, Instituut voor Kern- en Stralingsfysica, B-3001 Leuven, Belgium}

\author {P.~ Ring} 
\affiliation{Fakult{\"a}t f{\"u}r Physik, Technische Universit{\"a}t M{\"u}nchen, D-85748 Garching, Germany}

\author {H.~Nakada}
\affiliation{Department of Physics, Graduate School of Science, Chiba University, Yayoi-cho, I-33, Inage, Chiba 263-8522, Japan}

\author {A.N.~Andreyev}
\affiliation{Department of Physics, University of York, York, YO10 5DD, United Kingdom}
\affiliation{Advanced Science Research Center (ASRC), Japan Atomic Energy Agency (JAEA), Tokai-mura, Japan}

\author {P.~Van~Duppen}
\affiliation{KU~Leuven, Instituut voor Kern- en Stralingsfysica, B-3001 Leuven, Belgium}

\author {N.A.~Althubiti}
\affiliation{The University of Manchester, School of Physics and Astronomy, Oxford Road, M13 9PL Manchester, United Kingdom}
\affiliation{Physics Department, Faculty of Science, Jouf University, Aljouf, Saudi Arabia}

\author {B.~Andel}
\affiliation{KU~Leuven, Instituut voor Kern- en Stralingsfysica, B-3001 Leuven, Belgium}
\affiliation{Department of Nuclear Physics and Biophysics, Comenius University in Bratislava, 84248 Bratislava, Slovakia}


\author {D.~Atanasov}
\thanks{Present address: CERN, 1211, Geneva 23, Switzerland.}
\affiliation{Max-Planck-Institut f\"{u}r Kernphysik, Saupfercheckweg 1, 69117 Heidelberg, Germany}

\author {J.~Billowes}
\affiliation{The University of Manchester, School of Physics and Astronomy, Oxford Road, M13 9PL Manchester, United Kingdom}

\author {K.~Blaum}
\affiliation{Max-Planck-Institut f\"{u}r Kernphysik, Saupfercheckweg 1, 69117 Heidelberg, Germany}

\author {T.E.~Cocolios}
\affiliation{The University of Manchester, School of Physics and Astronomy, Oxford Road, M13 9PL Manchester, United Kingdom}
\affiliation{KU~Leuven, Instituut voor Kern- en Stralingsfysica, B-3001 Leuven, Belgium}

\author {J.G.~Cubiss}
\affiliation{Department of Physics, University of York, York, YO10 5DD, United Kingdom}


\author {G.J.~Farooq-Smith}
\affiliation{The University of Manchester, School of Physics and Astronomy, Oxford Road, M13 9PL Manchester, United Kingdom}
\affiliation{KU~Leuven, Instituut voor Kern- en Stralingsfysica, B-3001 Leuven, Belgium}

\author {D.V.~Fedorov}
\affiliation{Petersburg Nuclear Physics Institute, NRC Kurchatov Institute, Gatchina 188300, Russia}

\author {V.N.~Fedosseev}
\affiliation{CERN, CH-1211 Geneva 23, Switzerland}


\author {K.T.~Flanagan}
\affiliation{The University of Manchester, School of Physics and Astronomy, Oxford Road, M13 9PL Manchester, United Kingdom}
\affiliation{The Photon Science Institute, The University of Manchester, Manchester, M13 9PL, United Kingdom}

\author {L.P.~Gaffney}
\thanks{Present address: Department of Physics, University of Liverpool, Liverpool, L69 7ZE, United Kingdom}
\affiliation{KU~Leuven, Instituut voor Kern- en Stralingsfysica, B-3001 Leuven, Belgium}
\affiliation{School of Computing, Engineering, and Physical Sciences, University of the West of Scotland, Paisley PA1 2BE, United Kingdom}

\author {L.~Ghys}
\affiliation{KU~Leuven, Instituut voor Kern- en Stralingsfysica, B-3001 Leuven, Belgium}
\affiliation{Belgian Nuclear Research Center SCK$\bullet$CEN, Boeretang 200, B-2400 Mol, Belgium}

\author {M.~Huyse}
\affiliation{KU~Leuven, Instituut voor Kern- en Stralingsfysica, B-3001 Leuven, Belgium}

\author {S.~Kreim}
\affiliation{Max-Planck-Institut f\"{u}r Kernphysik, Saupfercheckweg 1, 69117 Heidelberg, Germany}
\affiliation{CERN, CH-1211 Geneva 23, Switzerland}

\author {D.~Lunney}
\thanks{Present address: Universit\'{e} Paris-Saclay, CNRS/IN2P3, IJCLab, 91405 Orsay, France.}
\affiliation{CSNSM-IN2P3, Universit\'{e} de Paris Sud, Orsay, France}

\author {K.M.~Lynch}
\affiliation{The University of Manchester, School of Physics and Astronomy, Oxford Road, M13 9PL Manchester, United Kingdom}
\affiliation{CERN, CH-1211 Geneva 23, Switzerland}

\author {V.~Manea}
\thanks{Present address: Universit\'{e} Paris-Saclay, CNRS/IN2P3, IJCLab, 91405 Orsay, France.}
\affiliation{Max-Planck-Institut f\"{u}r Kernphysik, Saupfercheckweg 1, 69117 Heidelberg, Germany}

\author {Y.~Martinez~Palenzuela}
\affiliation{KU~Leuven, Instituut voor Kern- en Stralingsfysica, B-3001 Leuven, Belgium}
\affiliation{CERN, CH-1211 Geneva 23, Switzerland}


\author {P.L.~Molkanov}
\affiliation{Petersburg Nuclear Physics Institute, NRC Kurchatov Institute, Gatchina 188300, Russia}

\author {M.~Rosenbusch}
\thanks{Present address: Wako Nuclear Science Center (WNSC), Institute of Particle and Nuclear Studies (IPNS), High Energy Accelerator Research Organization (KEK), Wako, Saitama 351-0198, Japan}
\affiliation{Universit\"{a}t Greifswald, Institut f\"{u}r Physik, 17487 Greifswald, Germany}

\author {R.E.~Rossel}
\affiliation{CERN, CH-1211 Geneva 23, Switzerland}
\affiliation{Institut f\"{u}r Physik, Johannes Gutenberg-Universit\"{a}t, D-55099 Mainz, Germany}

\author {S.~Rothe}
\affiliation{CERN, CH-1211 Geneva 23, Switzerland}

\author {L.~Schweikhard}
\affiliation{Universit\"{a}t Greifswald, Institut f\"{u}r Physik, 17487 Greifswald, Germany}

\author {M.D.~Seliverstov}
\affiliation{Petersburg Nuclear Physics Institute, NRC Kurchatov Institute, Gatchina 188300, Russia}

\author {P.~Spagnoletti}

\affiliation{School of Computing, Engineering, and Physical Sciences, University of the West of Scotland, Paisley PA1 2BE, United Kingdom}

\author {C.~Van~Beveren}
\affiliation{KU~Leuven, Instituut voor Kern- en Stralingsfysica, B-3001 Leuven, Belgium}

\author {M.~Veinhard}
\affiliation{CERN, CH-1211 Geneva 23, Switzerland}

\author{E.~Verstraelen}
\affiliation{KU~Leuven, Instituut voor Kern- en Stralingsfysica, B-3001 Leuven, Belgium}

\author{A.~Welker}
\affiliation{CERN, CH-1211 Geneva 23, Switzerland}
\affiliation{Institut f\"{u}r Kern- und Teilchenphysik, Technische Universit\"{a}t Dresden, Dresden 01069, Germany}

\author{K.~Wendt}
\affiliation{Institut f\"{u}r Physik, Johannes Gutenberg-Universit\"{a}t, D-55099 Mainz, Germany}

\author{F.~Wienholtz}
\thanks{Present address: Institut f\"{u}r Kernphysik, Technische Universit\"{a}t Darmstadt, 64289 Darmstadt, Germany.}
\affiliation{CERN, CH-1211 Geneva 23, Switzerland}
\affiliation{Universit\"{a}t Greifswald, Institut f\"{u}r Physik, 17487 Greifswald, Germany}

\author{R.N.~Wolf}
\thanks{Present address: ARC Centre of Excellence for Engineered Quantum Systems, School of Physics, The University of Sydney, NSW 2006, Australia.}
\affiliation{Max-Planck-Institut f\"{u}r Kernphysik, Saupfercheckweg 1, 69117 Heidelberg, Germany}
\affiliation{Universit\"{a}t Greifswald, Institut f\"{u}r Physik, 17487 Greifswald, Germany}

\author{A.~Zadvornaya}
\affiliation{KU~Leuven, Instituut voor Kern- en Stralingsfysica, B-3001 Leuven, Belgium}

\author{K. Zuber}
\affiliation{Institut f\"{u}r Kern- und Teilchenphysik, Technische Universit\"{a}t Dresden, Dresden 01069, Germany}

\date{\today}

\begin{abstract}
{

\noindent The mean-square charge radii of $^{207,208}$Hg ($Z=80, N=127,128$) have been studied for the first time and those of $^{202,203,206}$Hg ($N=122,123,126$) remeasured by the application of in-source resonance-ionization laser spectroscopy at ISOLDE (CERN). The characteristic \textit{kink} in the charge radii at the $N=126$ neutron shell closure has been revealed, providing the first information on its behavior below the $Z=82$ proton shell closure. A theoretical analysis has been performed within relativistic Hartree-Bogoliubov and non-relativistic Hartree-Fock-Bogoliubov approaches, considering both the new mercury results and existing lead data. Contrary to previous interpretations, it is demonstrated that both the kink at $N=126$ and the odd-even staggering (OES) in its vicinity can be described predominately at the mean-field level, and that pairing does not need to play a crucial role in their origin. A new OES mechanism is suggested, related to the staggering in the occupation of the different neutron orbitals in odd- and even-$A$ nuclei, facilitated by particle-vibration coupling for odd-$A$ nuclei.}

\end{abstract}


\maketitle

Experimental investigations of nuclear charge radii have revealed a rich abundance of regular patterns, abrupt changes and non-linear trends along isotopic chains across the nuclear chart~\cite{Angeli2009,Campbell2016}. Two near-universal features include kinks at neutron shell closures and odd-even staggering (OES), where an odd-$N$ isotope has a smaller charge radius than the average of its two even-$N$ neighbors~\cite{Angeli2009,Otten1989}. The commonality of these features indicates that their origin is general and independent of local microscopic phenomena. As such, measurements of kinks and OES provide a particularly stringent benchmark for nuclear theory~\cite{Tajima1993,Sharma1993,Fayans2000,Nakada2019,Gorges2019,deGroote2020}.
	
\begin{figure}[b]
	\includegraphics[width=1\linewidth]{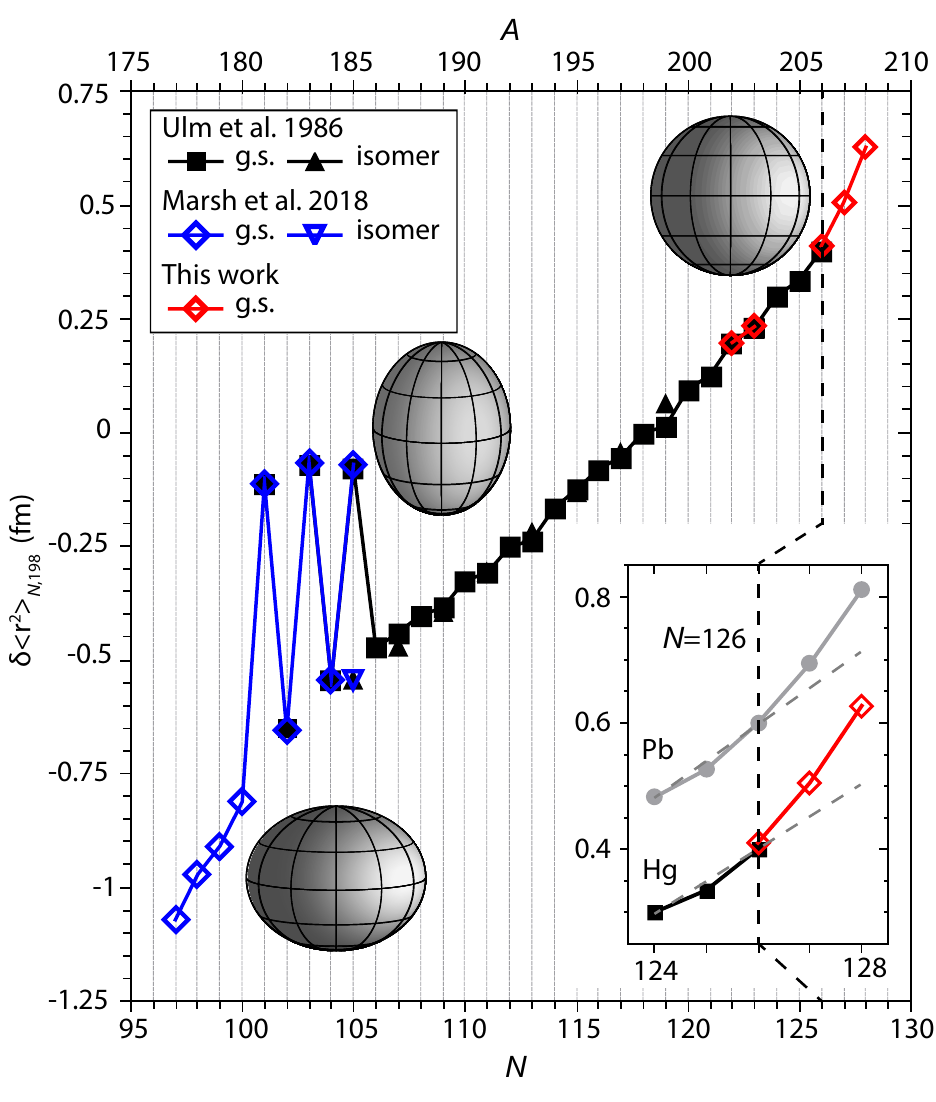}
	\caption{Systematics of the difference in mercury ground state (g.s.) and isomer mean-square charge radii. Data from the present work are shown by red symbols, earlier data are taken from~\cite{Ulm1986} (black) and \cite{Marsh2018} (blue). The inset highlights the kink at $N=126$ and the neighboring OES in both the mercury and lead~\cite{Anselment1986} isotopic chains. The lead isotopes are arbitrarily displaced from those of mercury for clarity. The dashed lines through $N=124, 126$ in the inset are added to highlight the kinks. Statistical uncertainties are smaller than the data points.
		\label{fig:Hg_NR}}
\end{figure}

Historically, the majority of the discussion pertaining to the effect of shell closures on charge radii focused on the kink in the lead isotopic chain across $N=126$~\cite{Tajima1993,Sharma1993,Fayans2000,Niksic2002,Goddard2013}, which is shown in the inset of Fig.~\ref{fig:Hg_NR}. A variety of theoretical approaches have been employed to investigate this kink. Relativistic mean field studies have described it with varying degrees of success~\cite{Sharma1993,Lalazissis1999}. By contrast, calculations in non-relativistic density functional theories (NR-DFTs) based on conventional functionals were unable to reproduce the kink~\cite{Tajima1993,Gorges2019}. The differences between the descriptions in these theoretical frameworks are related to the relative occupations of the neutron $\nu 1i_{11/2}$ and $\nu 2g_{9/2}$ orbitals (located above the $N=126$ shell closure), with the magnitude of the kink being driven by the occupation of the $\nu 1i_{11/2}$ orbital~\cite{Goddard2013}. The extension of non-relativistic functionals, by the addition of gradient terms into the pairing interaction, has been demonstrated as a possible way to improve their description of OES and the kink~\cite{Fayans2000,Gorges2019}. An alternative approach is the inclusion of a density dependence in the spin-orbit interaction~\cite{Nakada2019,Nakada2015,Nakada2015a}, derived from the chiral three-nucleon interaction by Kohno~\cite{Kohno2012}.

The optimal method of quantifying the kink at $N=126$ considers isotopes with $N=124, 126, 128$, to avoid contributions from either OES or possible deformation in isotopes further from the shell closure. However, in contrast to the regions near $N=50$ and $N=82$~\cite{Angeli2009}, there are limited experimental data on charge radii behavior across $N=126$ (specifically for $N=128$), with corresponding measurements available only for $Z=82$~\cite{Anselment1986} and $Z=83$~\cite{Barzakh2018}. 

In this Letter we report the first study of charge radii across $N=126$ in the mercury $(Z=80)$ isotopic chain, thus enabling the $Z$-dependence of the kink at $N=126$ to be probed and providing the first benchmark for theory in the region below the $Z=82$ proton shell closure. These new data motivated us to undertake a comparative theoretical investigation of the kinks and OES in lead and mercury charge radii across $N=126$. By applying both spherical relativistic Hartree-Bogoliubov (RHB)~\cite{Vretenar2005} and spherical non-relativistic Hartree-Fock-Bogoliubov (NR-HFB)~\cite{Nakada2020} approaches, we explored whether an alternative explanation of the kink and OES is possible. This was also the first study of OES within a relativistic framework. 

The experimental data originates from the same measurement campaign as described in Refs.~\cite{Marsh2018,Sels2019}, where neutron-deficient mercury isotopes were also studied (Fig.~\ref{fig:Hg_NR}). Therefore, only a brief experimental overview is included here. Mercury isotopes were produced at the CERN-ISOLDE facility~\cite{Catherall2017} by impinging a 1.4-GeV proton beam from the Proton Synchrotron Booster on to a molten-lead target. The neutral reaction products effused 
into the anode cavity of a Versatile Arc Discharge and Laser Ion Source (VADLIS)~\cite{DayGoodacre2016b}. Laser light from the ISOLDE Resonance Ionization Laser Ion Source (RILIS) complex~\cite{Fedosseev2017} was used to excite three sequential atomic transitions, for the resonance ionization of the mercury isotopes~\cite{DayGoodacre2016c}. The photo-ions were extracted and mass separated by the ISOLDE general purpose separator and then directed to either a Faraday cup for direct photo-ion detection, or to the ISOLTRAP Paul trap~\cite{Herfurth2001} and multi-reflection time-of-flight mass spectrometer (MR-ToF MS)~\cite{Wolf2013} for single-ion counting and discrimination from isobaric contamination. 

\begin{figure}
	\includegraphics[width=1\linewidth]{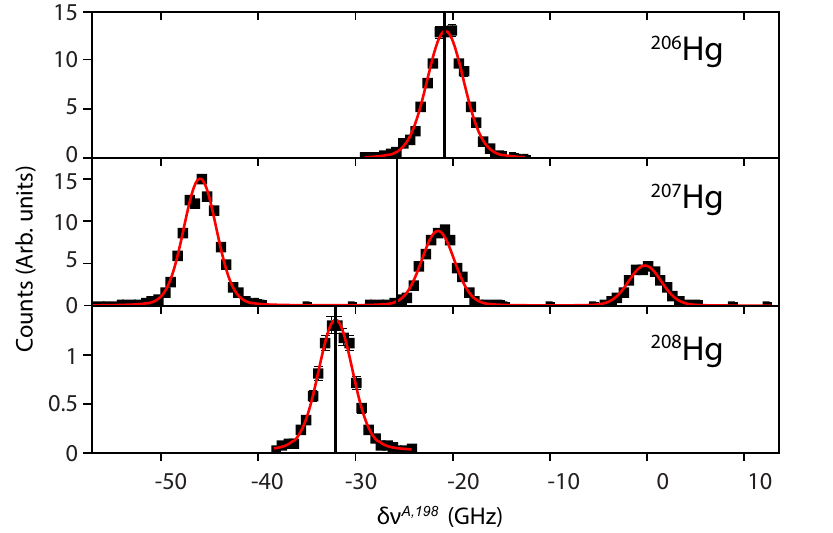}
	\caption {Sample hyperfine spectra for $^{206-208}$Hg, with the photo-ion rate measured using the MR-ToF MS. The centroids are indicated with black lines, red lines represent fitting with Voigt profiles.} 
	\label{fig:laser_scans}
\end{figure}

The first of the three atomic transitions ($5d^{10}6s^2$~$^1S_0$~$\rightarrow$~$5d^{10}6s6p$~$^3P_1^{\circ}$, 253.65~nm) was probed by scanning a frequency-tripled titanium-sapphire laser (full width at half maximum bandwidth $<$1~GHz before tripling)~\cite{Rothe2013}. Isotope shifts (IS), $\delta \nu ^{A,A'}=\nu ^{A}-\nu ^{A'}$,  in the frequency of this transition were measured for mercury nuclei with $A=202,203,206,207,208$ relative to the stable reference isotope with $A'=198$. Sample spectra are presented in Fig.~\ref{fig:laser_scans}. Details of the scanning and fitting procedures can be found in Refs.~\cite{Rossel2015,Sels2019} and Refs.~\cite{Sels2019,Seliverstov2014}, respectively, with further information on the data analysis in Refs.~\cite{DayGoodacre2016,Sels2018}. The relative changes in the mean-square charge radii, $\delta\langle r^{2}\rangle^{A,A'}$, were extracted from the measured $\delta \nu ^{A,A'}$ values via standard methods described in the supplemental material~\cite{Supmat}.
 
The extracted $\delta \nu ^{A,A'}$ and $\delta\langle r^{2}\rangle^{A,A'}$ values are presented in Table~\ref{tab:Exp_results} and the $\delta\langle r^{2}\rangle^{A,A'}$ data are plotted in Fig.~\ref{fig:Hg_NR}. There is a visible kink at $N=126$, with a magnitude similar to that in the lead isotopic chain. The IS for $^{202,203}$Hg, and the remeasured neutron-deficient isotopes~\cite{Marsh2018,Sels2019} from the same experimental campaign, are in good agreement with literature. For $^{206}$Hg there is a $\approx$ 500-MHz discrepancy between this work and the thesis value of Ref.~\cite{Dabkiewicz} cited in~\cite{Ulm1986}, which is discussed in Ref.~\cite{DayGoodacre2016}. Given the agreement of the results from the present experimental campaign with the other literature data, the ${\delta \nu^{206, 198}}$ value from this work is used for the following discussion.

\begin{table}[b]
	\caption{\label{tab:Exp_results} $\delta \nu^{A, 198}$ values in the 253.65~nm line from this work and literature. $\delta \langle r^2\rangle ^{A, 198}$ values are calculated as described in the supplemental material~\cite{Supmat}. Statistical uncertainties are shown in parentheses and systematic uncertainties	in curly brackets.}
\begin{ruledtabular}
	\begin{tabular}{ccccc}
		\textrm{Isotope}&
		\textrm{$I^{\pi}$}&
		\textrm{$\delta \nu^{A, 198}$}&
		\textrm{$\delta \langle r^2\rangle ^{A, 198}$}&
		\textrm{Ref.} \\

		A & & (MHz) & (fm$^2$) & \\
		\colrule
		202&0 &-10 100(180) &0.197(3)\{14\}	&This work\\
		&& -10 102.4(42) &0.1973(1)\{138\}&\cite{Ulm1986}\\ 
		203&5/2$^-$& -11 870(200) &0.232(4)\{16\}& This work\\
		&& -11 750(180) &	0.2295(35)	\{161\} &\cite{Ulm1986}\\ 
		206&0& -20 930(160) & 0.409(3)\{29\} & This work\\
		&&  -20 420(80) & 0.3986(16)\{280\}& \cite{Ulm1986,Dabkiewicz}\\
		207&(9/2$^-$)&-25 790(190)&0.503(4)\{35\}&This work\\
		208&&-32 030(160)&0.624(3)\{44\}&This work\\ 
	\end{tabular}
\end{ruledtabular}
\end{table}

To interpret the data, a new RHB code was developed which enables the blocking of selected single-particle orbitals and allows for fully self-consistent calculations of the ground and excited states in odd-$A$ nuclei. A separable version of the Gogny pairing is used~\cite{Tian2009}, with the pairing strength of Ref.~\cite{Agbemava2014}. The NL3*, DD-PC1, DD-ME2 and DD-ME$\delta$ covariant energy density functionals (CEDF) were employed, the global performance of which was tested in Ref.~\cite{Agbemava2014}. The functionals achieved a comparable description of the kink and the OES, thus only the DD-ME2 results are discussed below. The results for the other functionals will be included in a follow-up paper~\cite{DayGoodacre2021}. 

The NR-HFB calculations were performed assuming spherical symmetry with the semi-realistic M3Y-P6a interaction, the spin-orbit properties of which were modified~\cite{Nakada2015} to improve the description of the charge radii of proton-magic nuclei~\cite{Nakada2019, Nakada2015a, Nakada2015}. We applied it here for the first time to the mercury isotopic chain. For $N \leq 126$, isotopes with $\langle\beta_2^2\rangle^{1/2} < 0.1$ were considered, where $\langle\beta_2^2\rangle$ is the mean-square deformation
deduced from $\delta \langle r^2 \rangle$ using the droplet model~\cite{Otten1989, Berdichevsky1985}. This restriction corresponds to $N\geq 116$ and $N\geq 121$ for lead and mercury isotopes, respectively.

The importance of a simultaneous agreement of energetic and geometric nuclear observables in such investigations has been highlighted previously~\cite{Fayans2000}. Thus, in addition to calculating the charge radii, we also checked the quality of the binding energy description. 
There is a good agreement between the DD-ME2 and experimental binding energies for both lead and mercury isotopes~\cite{Audi2017}. The rms deviation is 1.3~MeV and 1.1~MeV for $^{198-214}$Pb and $^{201-208}$Hg, respectively, a comparable performance to the widely used NR-DFT functional UNEDF1~\cite{Kortelainen2012} (1.4~MeV and 1.0~MeV for lead and mercury, respectively~\cite{ME}). The binding and separation energy descriptions of all of the employed CEDFs will be discussed in detail in Ref.~\cite{DayGoodacre2021}.

Two different procedures labeled as ``LES" and ``EGS" are used for the blocking in odd-$A$ nuclei and the results of respective calculations are labeled by the ``Functional-Procedure" labels (for example, DD-ME2-EGS). In the LES procedure, the lowest in energy configuration is used: this is similar to all earlier calculations of OES~\cite{Fayans2000,Reinhard2017}. In the EGS procedure, the configuration with the spin and parity of the blocked state corresponding to those of the experimental ground state is employed, although it is not necessarily the lowest in energy. For example, in the RHB(DD-ME2) calculations the EGS configurations with a blocked $\nu 2g_{9/2}$ state are located at excitation energies of 137~keV, 122~keV and 96~keV above the ground state configurations with a blocked $\nu 1i_{11/2}$ state in $^{209,211,213}$Pb. At first glance, this contradicts experimental findings  that the ground state is based on the $\nu 2g_{9/2}$ orbital in odd-$A$ lead isotopes with $N>126$. However, particle-vibration coupling (PVC) lowers the energy of this state below that of the $\nu 1i_{11/2}$ one (see Fig. 5 in Ref.~\cite{Litvinova2011}) so that it becomes the lowest in energy in the PVC calculations. Note that PVC significantly improves the accuracy of the description of the energies of experimental states in model calculations ~\cite{Litvinova2011,Afanasjev2015}. However, it is neglected in the present study since its impact on charge radii is still an open theoretical question.

\begin{figure}[b]
	\includegraphics[width=1\linewidth]{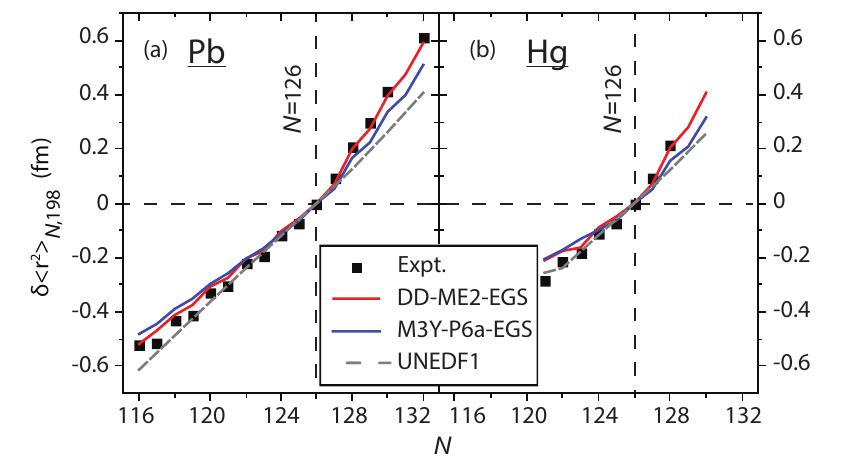}%
	\caption{\label{fig:CR_Comp} $\delta \langle r^2\rangle ^{A, A'}$ of lead (a) and mercury (b) isotopes relative to $^{208}$Pb and $^{206}$Hg ($N=126$), respectively. Experimental mercury data: this work and~\cite{Ulm1986}, experimental lead data:~\cite{Anselment1986}, the statistical uncertainties are smaller than the data points. RHB(DD-ME2) results: this work, NR-HFB(M3Y-P6a) results: this work (mercury) and~\cite{Nakada2015a} (lead), NR-HFB(UNEDF1) results:~\cite{ME}. }
\end{figure}

The results of the RHB and NR-HFB calculations are presented in Fig.~\ref{fig:CR_Comp}, together with the experimental results for the lead and mercury chains. In both cases, the kink at $N=126$ is visibly better reproduced in the RHB (DD-ME2) calculations.
To facilitate a quantitative comparison of the experimental and theoretical results, two indicators are employed. OES is quantified considering the isotope's nearest neighbors via the commonly used three-point indicator
\begin{equation}
\Delta \langle r^2 \rangle^{(3)}(A) = \frac{1}{2} \left[ \langle r^2(A-1) \rangle +
\langle r^2(A+1) \rangle - 2  \langle r^2(A) \rangle \right].
\label{eqn:3pi}
\end{equation}

\noindent To quantify the shell effect at $N=126$, the kink indicator of Ref.~\cite{Gorges2019} is used which considers the isotope's next-to-nearest neighbors and is defined as

\begin{equation}
\Delta R^{(3)}(A) = \frac{1}{2} \left[ R(A-2)  +
R(A+2)  - 2  R(A)  \right],
\label{eqn:4pi}
\end{equation}

\noindent where $R(A)=\langle r^2 \rangle^{1/2}(A)$ is the charge radius of the isotope with mass $A$ of the element under consideration. Note that the kink indicator is independent of the blocking procedure in odd-$A$ nuclei and therefore we omit the corresponding 
specifications (LES or EGS) in the discussion of the kink.

$\Delta \langle r^2 \rangle^{(3)}(A)$ and $\Delta R^{(3)}(A)$ values calculated from the experimental results and theoretical calculations for both lead and mercury are presented in Fig.~\ref{fig:3pi}, and the $\Delta R^{(3)}(A)$ values are listed in the supplemental material~\cite{Supmat}. It is evident in Figs.~\ref{fig:3pi}~(a) and~\ref{fig:3pi}~(b) that the magnitude of the kinks in the isotopic chains are comparable, suggesting that the kink at $N=126$ is broadly insensitive to the change of the occupied proton states when crossing $Z=82$ ($\pi 2d_{3/2}$ in mercury and $\pi 3s_{1/2}$ in lead). In addition, the RHB(DD-ME2) calculations best reproduce the kink, while the NR-HFB(M3Y-P6a) and NR-HFB(Fy($\Delta$r)~\cite{Gorges2019}) results underestimate and overestimate its magnitude, respectively. It is worth noting that both the RHB(DD-ME2) and NR-HFB(M3Y-P6a) approaches are reasonably successful in the reproduction of the absolute charge radius values for $^{206}$Hg and $^{208}$Pb, the details are included in the supplemental material~\cite{Supmat}.

In both approaches, the OES is best reproduced if the EGS procedure is applied (see Figs.~\ref{fig:3pi} (c), (d) and Fig.~\ref{fig:pnp}). If the LES procedure is applied, the experimental OES is significantly underestimated for all nuclei under study in the RHB calculations and for $N<126$ nuclei in the NR-HFB calculations. Note that for simplicity we show only NR-HFB results with both procedures
in  Fig.~\ref{fig:3pi} (c) and (d).

For a better understanding of the underlying mechanisms of both the kink and OES, we also performed RHB calculations without pairing for lead isotopes. The labels identifying such results contain ``np". Significantly, a kink is still present in the results as depicted in Fig.~\ref{fig:3pi}~(a), due to the occupation of the $\nu 1i_{11/2}$ orbital. This indicates a mechanism alternative to the one based on gradient terms in pairing interactions~\cite{Fayans2000, Gorges2019}. 
  
\begin{figure}[b]%
	\includegraphics[width=1\linewidth]{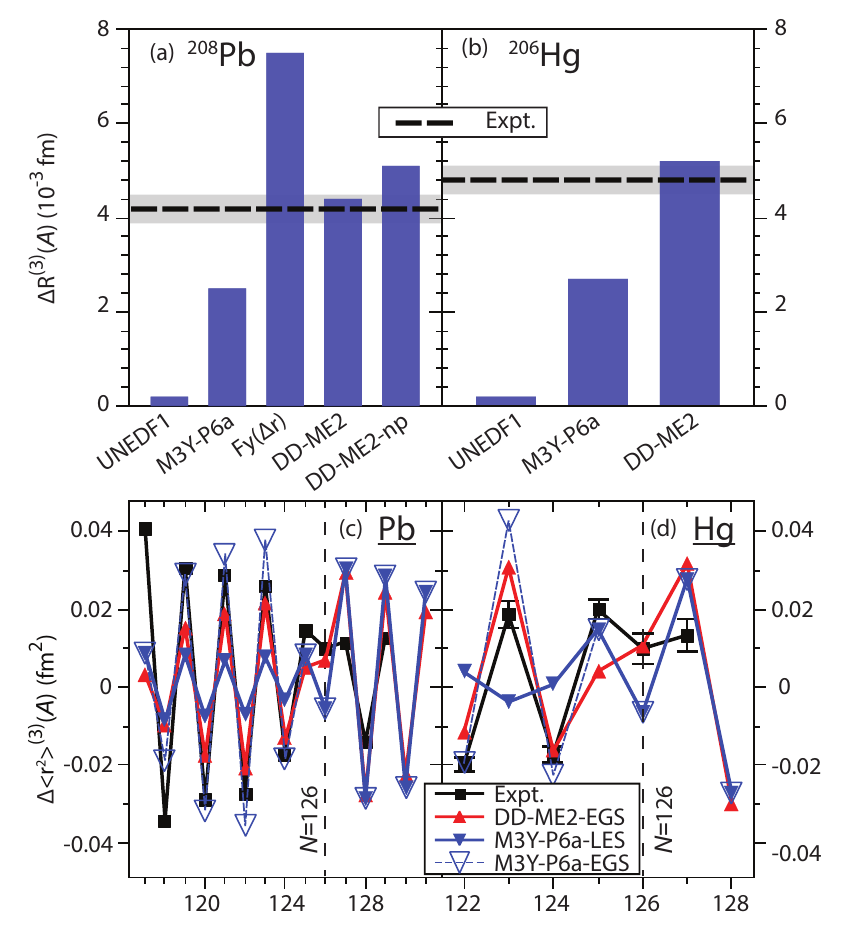}%
	\caption{Comparison of experimental and theoretical $\Delta R^{(3)}(A)$ and $\Delta \langle r^2 \rangle^{(3)}(A)$ 
	values for isotopes of lead, (a) and (c), respectively, and mercury (b) and (d), respectively. Experimental values are taken from this work and from Ref.~\cite{Ulm1986} (mercury) and Ref.~\cite{Anselment1986} (lead). The RHB(DD-ME2) and NR-HFB(M3Y-P6a) results are obtained in this work and the NR-HFB results with Fy($\Delta r$) and UNEDF1 are taken from Refs.~\cite{Gorges2019} and~\cite{ME}, respectively. Experimental uncertainty is depicted as translucent gray bars in (a) and (b), and as error bars in (c) and (d).}
	\label{fig:3pi}%
\end{figure}%

The RHB results with and without pairing are compared via $\Delta \langle r^2 \rangle^{(3)}(A)$ in Fig.~\ref{fig:pnp}. OES appears in these calculations (the curves labeled as  ``DD-ME2-EGS"  and  ``DD-ME2-np-EGS") under the condition that, in odd-$A$ nuclei, the EGS procedure  is used. One can see that the inclusion of pairing somewhat reduces this effect. However, OES is mostly absent if the LES procedure is used in odd-$A$ nuclei.

\begin{figure}%
	\includegraphics[width=1\linewidth]{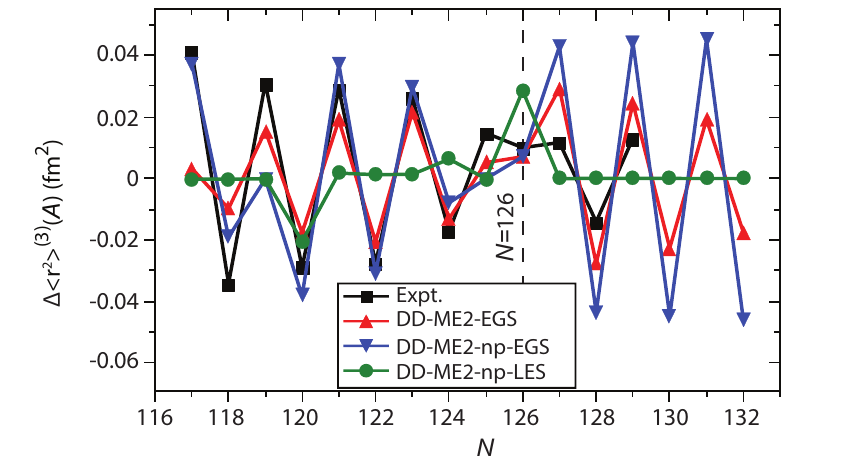}%
	\caption{Comparison of experimental and theoretical $\Delta \langle r^2 \rangle^{(3)}(A)$ values for lead isotopes. Experimental data from~\cite{Anselment1986}, theoretical results are from this work. See text for details.}	
	\label{fig:pnp}%
\end{figure}%

Let us consider the lead isotopes with $N\geq 126$ for a more detailed discussion of the origin of OES in the calculations without pairing. By designating the ground state of $^{208}$Pb as a ``core" and noting that PVC lowers the energy of the $\nu 2g_{9/2}$ state below $\nu 1i_{11/2}$ in odd-$A$ nuclei~\cite{Litvinova2011}, the sequence of the ground states in the $N\geq 126$ nuclei can be described as ``core" ($^{208}$Pb), ``core"$\otimes \nu (2g_{9/2})^1$ ($^{209}$Pb), ``core"$\otimes \nu (2i_{11/2})^2$  ($^{210}$Pb), ``core"$\otimes \nu (2i_{11/2})^2 (2g_{9/2})^1$ ($^{211}$Pb) and so on in the relativistic calculations without pairing. The $\nu1i_{11/2}$ orbital has a smaller rms-radius than the $\nu 2g_{9/2}$ orbital. However, because of the isovector nature of nuclear forces its occupation leads to a larger charge radius as compared with the occupation of $\nu 2g_{9/2}$ orbital. Thus,  the staggering in their  occupation between odd and even isotopes results in the OES seen in Fig.~\ref{fig:pnp}.

On the contrary, in the majority of conventional non-relativistic functionals, the  $\nu 2g_{9/2}$ orbital is lower in energy than the  $\nu 1i_{11/2}$ orbital. This is in agreement with experimental data on the structure of the ground states in odd-mass nuclei, but creates a problem in the description of the kinks. In addition, in calculations with and without pairing this leads to the sole or predominant occupation of the $\nu 2g_{9/2}$ state in even-even and odd-even nuclei with $N>126$ and thus to a negligible or comparatively small OES. To address this, several prescriptions have been suggested over the years to increase the occupation of the $\nu 1i_{11/2}$ orbital in the $N>126$~lead nuclei. One approach includes a modification of the spin-orbit interaction leading either to the inversion of the relative energies of these two states or to their proximity in energy~\cite{Sharma1995,Reinhard1995,Stone2007,Bender2003,Erler2011,Goddard2013}. The NR-HFB results with M3Y-P6a shown in Figs.~\ref{fig:CR_Comp} and~\ref{fig:3pi} are also based on a modification of the spin-orbit interaction, with the inclusion of a density dependent term in the spin-orbit channel.  
Alternatively, the so-called Fayans functionals employ a specific form of the pairing interaction containing a gradient term~\cite{Fayans1994,Fayans2000,Reinhard2017,Gorges2019}. Although this improves the general description of experimental data, discrepancies between theory and experiment still exist in the lead and tin isotopic chains \cite{Gorges2019}. Moreover, pairing becomes a dominant contributor to the kink and OES~\cite{Gorges2019}. 

The present RHB interpretation of the kinks and OES differs from that suggested in~\cite{Gorges2019}, which is based on non-relativistic Skyrme and Fayans functionals. In the RHB approach, the kink and OES are already present in the calculations without pairing. Thus, the evolution of charge radii with neutron number depends significantly on the mean-field properties. Pairing acts only as an additional tool which mixes different configurations and somewhat softens the evolution of charge radii as a function of neutron number.

In conclusion, the determination of the $\delta \langle r^2\rangle ^{A, A'}$ of $^{207,208}$Hg has revealed a kink at $N=126$ in the mercury nuclear charge radii systematics, with a magnitude comparable to that in the lead isotopic chain. These new data have been analyzed via both RHB and NR-HFB approaches, together with the traditional magic-$Z$ theoretical benchmark of the lead isotopic chain. We demonstrate that the kinks at the $N=126$ shell closure and the OES in the vicinity, are currently best described in the RHB approach without any readjustment of the parameters defined in Ref.~\cite{DD-ME2}. According to the RHB calculations, the kink at $N=126$ in $\delta \langle r^2\rangle ^{A, A'}$ originates from the occupation of the $\nu 1i_{11/2}$ orbital located above the $N=126$ shell gap. A new mechanism for OES is suggested, related to the staggering in the occupation of neutron orbitals between odd and even isotopes and facilitated by PVC in odd-mass nuclei. Thus, contrary to previous interpretations, it is determined that both the kink and OES in charge radii can be defined predominantly in the particle-hole channel.

\begin{acknowledgments}
\label{sec:Acknowledgements}This project has received funding through the European Union's Seventh Framework Programme for Research and Technological Development under grant agreements 267194 (COFUND), 262010 (ENSAR), 289191 (LA$^{3}$NET) and 267216 (PEGASUS). This project  has received funding from the European Union’s Horizon 2020 research and innovation programme grant agreement No 654002. This material is based upon work supported by the US Department of Energy, Office of Science, Office of Nuclear Physics under Award No. DE-SC0013037, and by the DFG cluster of excellence ``Origin and Structure of the Universe" (www.universe-cluster.de). S. S. acknowledges a SB PhD grant from the former Belgian Agency for Innovation by Science and Technology (IWT), now incorporated in FWO-Vlaanderen. This work was supported by the RFBR according to the research project N 19-02-00005; the ERC Consolidator Grant No.~648381; the IUAP-Belgian State Science Policy (BRIX network P7/12), FWO-Vlaanderen (Belgium) and GOA’s 10/010 and 10/05 and starting grant STG 15/031 from KU Leuven; the Science and Technology Facilities Council Consolidated Grant Nos. ST/F012071/1 and ST/P003885/1, Continuation Grant No.~ST/J000159/1 and Ernest Rutherford Grant No.~ST/L002868/1; the Slovak Research and Development Agency, contract No. APVV-14-0524; the BMBF (German Federal Ministry for Education and Research) grants Nos. 05P12HGCI1, 05P15HGCIA and 05P18HGCIA.

\end{acknowledgments}

\subsection{}
\subsubsection{}

\bibliography{Hg_NR_L_wt_m_red}
\pagebreak
\appendix



\end{document}